\documentclass[pra,aps,superscriptaddress,twocolumn,showpacs]{revtex4}

\usepackage{graphicx}
\usepackage{amsfonts}
\usepackage{amssymb}
\usepackage{amsmath}
\usepackage{bm}
\usepackage{color}
\usepackage[normalem]{ulem}
\newcommand{\be}{\begin{equation}}
\newcommand{\ee}{\end{equation}}
\newcommand{\bea}{\begin{eqnarray}}
\newcommand{\eea}{\end{eqnarray}}
\newcommand{\Eq}[1]{Eq.\,(\ref{#1})}
\newcommand{\Fig}[1]{Fig.\,\ref{#1}}
\newcommand{\Sec}[1]{Sec.\,\ref{#1}}
\newcommand{\Cite}[1]{~[\onlinecite{#1}]} 
\newcommand{\Onlinecite}[1]{Ref.\,[\onlinecite{#1}]} 

\renewcommand{\Re}{\mathrm{Re}}
\newcommand{\E}{\mathcal{E}}
\newcommand{\tE}{\widetilde{\mathcal{E}}}

\DeclareMathOperator*{\SI}{\sum\hspace{-13pt}\int}

\newcommand{\App}[1]{Appendix\,\ref{#1}}

\begin{document}
\title{Resonant-state expansion of light propagation in non-uniform waveguides }
\author{S.\,V. Lobanov}\email{LobanovS@cardiff.ac.uk}
\author{G. Zoriniants}
\author{W. Langbein}
\author{E.\,A. Muljarov}\email{egor.muljarov@astro.cf.ac.uk}
\affiliation{School of Physics and Astronomy, Cardiff University, Cardiff CF24 3AA,
United Kingdom}
\begin{abstract}
A new rigorous approach for precise and efficient calculation of light propagation along non-uniform waveguides is presented. Resonant states of a uniform waveguide, which satisfy outgoing-wave boundary conditions, form a natural basis for expansion of the local electromagnetic field. Using such an expansion at fixed frequency, we convert the wave equation for light propagation in a non-uniform waveguide into an ordinary second-order matrix differential equation for the expansion coefficients depending on the coordinate along the waveguide.  We illustrate the method on several examples of non-uniform planar waveguides and evaluate its efficiency compared to the aperiodic Fourier modal method and the finite element method, showing improvements of one to four orders of magnitude. A similar improvement can be expected also for applications in other fields of physics showing wave phenomena, such as acoustics and quantum mechanics.
\end{abstract}
\pacs{03.50.De, 42.25.-p, 03.65.Nk}
\date{\today}
\maketitle

\section{Introduction}\label{sec:intro}

Uniform optical waveguides (WGs), such as a dielectric slab in vacuum, are translationally invariant systems which support bound states of light called WG modes~\cite{Marcuse91}.
These modes present a small, though significant subgroup of a larger class of resonant states (RSs) of an optical system, among which there are also unbound solutions, such as Fabry-Perot (FP) and anti-WG modes~\cite{Armitage14}. Formally, RSs are the eigenmodes of an open optical system, which satisfy either incoming or outgoing wave boundary conditions (BCs), and describe, with mathematical rigor, optical resonances of different linewidth which exist in the system. WG modes correspond to infinitely narrow resonances, representing stable propagating waves.

\begin{figure}
	\includegraphics[width=\linewidth]{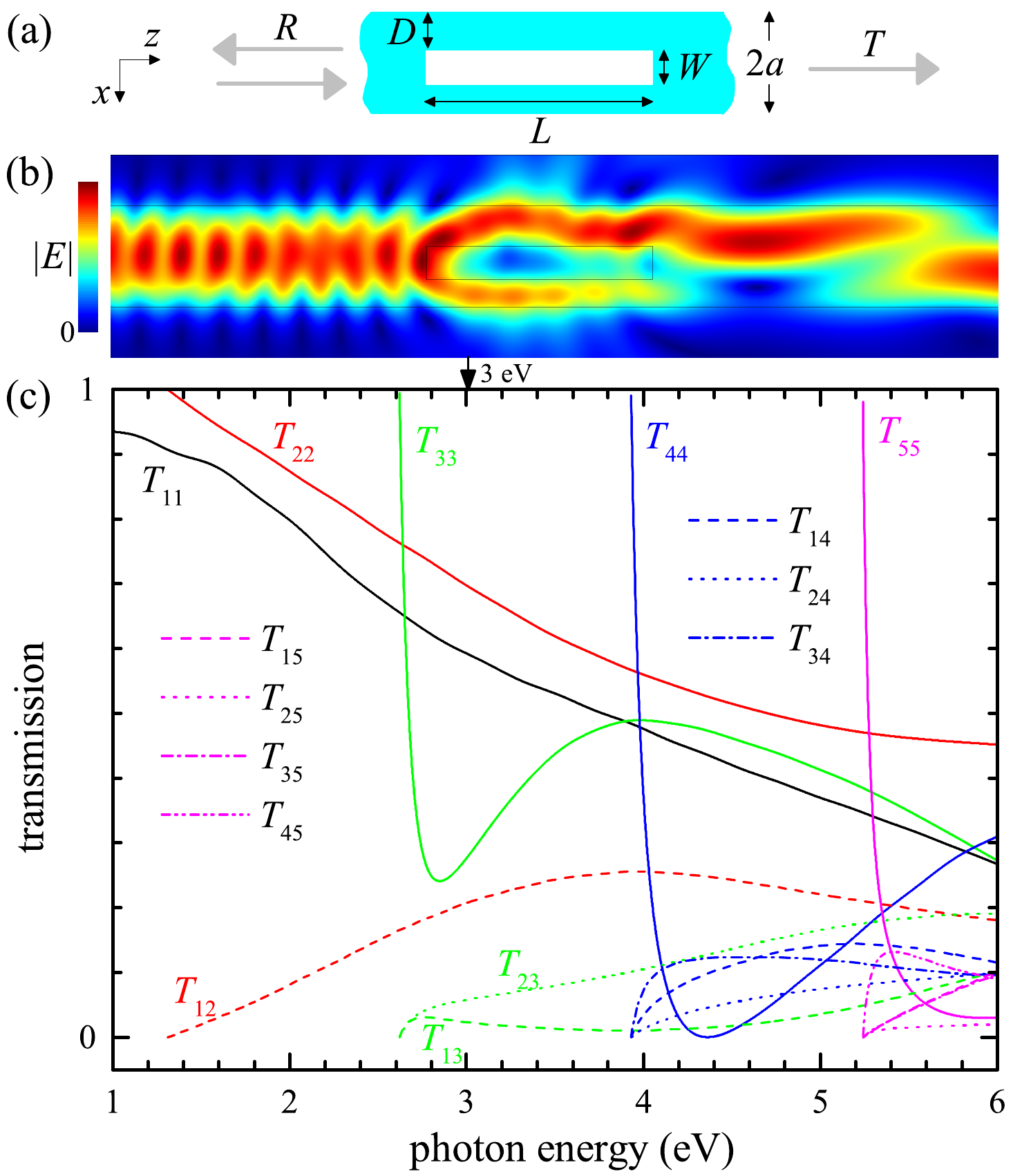}
	\caption{(a) Sketch of the scattering geometry and the considered planar dielectric waveguide with a rectangular hole. (b) Electric field amplitude for excitation with WG mode 1 at $\hbar\omega=3$\,eV from the left, on a linear color scale as given, overlaid with the WG outline. (c) Relative power transmission $T_{ij}$, from incoming left WG mode $j$ to outgoing right WG mode $i$, as function of the photon energy $\hbar\omega$.}
	\label{fig:Spectra}
\end{figure}

Non-uniform WGs have a varying cross-section along the main propagation direction. An electromagnetic (EM) wave, initially excited in a WG mode of a uniform region,  is scattered on WG inhomogeneities and can thus be transferred into other WG modes, see an example in \Fig{fig:Spectra}. However, some part of the EM energy leaks out of the system, an effect which is often treated using a continuum of radiation modes~\cite{Marcuse91}. This treatment does not make use of the natural unbound RSs, and is numerically costly, as an artificial discretization of the continuum has to be introduced. Using instead the contributions to the EM field of all RSs, including the unbound ones, the role of the radiation continuum can be minimized or even fully eliminated. This is achieved by modifying the contour of integration over the continuum in the complex wave number plane, as was suggested e.g. in \Onlinecite{Tamir63,Shevchenko71}, or by making a transformation from the frequency to the wave number plane \Cite{Armitage14}.

Several numerical methods of computational electrodynamics are presently employed for modelling of light propagation in non-uniform WGs. One popular approach is the aperiodic Fourier modal method (a-FMM)\Cite{Lalanne00,Silberstein01,Hugonin05,Pisarenco10}, a generalization of the standard FMM\Cite{Moharam83,Moharam95,Tikhodeev02}, which allows to treat an open WG by introducing an artificial periodicity and a perfectly matched layer (PML)\Cite{Berenger94,Silberstein01}. Other approaches include the finite difference in time domain method\Cite{Yee66,FDTD93} or the finite element method\Cite{FEM12}, both using a PML to mimic the outgoing wave BCs. Furthermore, the multimode moment method\Cite{Bhattacharyya94}, the mode matching technique\Cite{Eleftheriades94}, and the eigenmode expansion method\Cite{Gallagher03} use the eigenmodes of homogeneous WG regions explicitly, expanding the EM field in each uniform region into its own WG and radiation modes and then matching the field at inhomogeneities. Typically such expansions are limited to only WG modes\Cite{Katsenelenbaum98,Hardy94} neglecting the radiation continuum, which simplifies the calculation but results in systematic errors which are hard to control.

In this paper, we present the waveguide resonant-state expansion (WG-RSE), a new general method, based on the concept of RSs, for calculating light propagation in WGs with varying cross-section. Similar to some of the methods mentioned above, we expand the EM field into a complete set of eigenmodes of a homogeneous WG. However, we introduce two major advances: (i) we minimize the contribution of the radiation continuum by replacing it with the discrete unbound RSs, and (ii) we expand the field in all regions of the WG into the same basis RSs, in this way automatically fulfilling the mode matching conditions, which enables also treating continuous inhomogeneities. Both features are unique to our approach and make it orders of magnitude more efficient than other available methods.

\section{Formulation of WG-RSE}

The formalism of RSs has been recently applied to a uniform planar WG, and all types of RSs, including WG, anti-WG, and FP modes were calculated for an infinitely extended dielectric slab surrounded by vacuum\Cite{Armitage14}. It has also been shown that in spite of their exponential growth  outside the WG, unbound RSs naturally discretize the continuum of radiation modes and are suited for expansion of the EM field inside the WG.

Based on the concept of RSs, a rigorous approach in electrodynamics called resonant-state expansion (RSE) has recently been developed~\cite{Muljarov10}, enabling accurate calculation of RSs in photonic systems~\cite{Doost12,Doost13,Doost14,Muljarov16,Weiss16}. The RSE calculates RSs of a given optical system using RSs of a basis system which is typically analytically treatable, as a basis for expansion, and maps Maxwell's wave equation onto a linear matrix eigenvalue problem. This approach has been applied to uniform WGs~\cite{Armitage14} for the case of a fixed real in-plane propagation wavevector. For the description of propagation along a waveguide, we consider here instead RSs for a fixed real frequency, having in general complex in-plane wavevectors, and use them to formulate a fixed-frequency RSE for homogeneous parts of WGs. To treat non-uniform WGs, we expand the EM field into the basis RSs, with expansion coefficients varying along the WG. The propagation along the WG is then simply expressed by an ordinary second-order matrix differential equation for the expansion coefficients, which is the main result of the WG-RSE.

Let us now develop the general formalism of the WG-RSE, using as example a non-uniform planar WG in vacuum, translationally invariant in $y$-direction and having a varying cross-section in the $z$-direction, as sketched in \Fig{fig:Spectra}(a). The light propagation in this system is described by Maxwell's equations, which are reduced to a 2D scalar wave equation
\be \left(\frac{\partial^2}{\partial x^2}+\frac{\partial^2}{\partial z^2}+\omega^2\varepsilon(x,z)\right)\E(x,z)=0
\label{WE-2D} \ee
in the case of a TE-polarized electric field of the form ${\bf E}({\bf r},t)=\hat{\bf y} e^{-i\omega t}{\cal E}(x,z)$, oscillating with a fixed frequency $\omega>0$, 
where $\varepsilon(x,z)$ is the permittivity of the WG, $\hat{\bf y}$ is the unit vector along the $y$-axis, and the speed of light in vacuum $c=1$ is used.

To solve \Eq{WE-2D} we introduce a basis waveguide (BWG) which is defined as an infinitely extended homogeneous dielectric slab in vacuum, having a constant permittivity $\epsilon$ and a thickness $2a$ including all variations of the permittivity $\varepsilon(x,z)$ along the non-uniform WG.
The solution of \Eq{WE-2D} outside the BWG is known to be a superposition of plane waves $\exp(ipz\pm ikx)$ with
real wavenumbers $p$, and $k=\sqrt{\omega^2-p^2}$ positive real for $|p|<\omega$ (outgoing propagating waves) and positive imaginary for $|p|>\omega$ (evanescent waves). This allows us, using Maxwell's BCs, to reduce the problem \Eq{WE-2D} to the BWG region $|x|\leqslant a$ only, supplemented by the two BCs
\begin{equation}
\left(i\frac{d}{d x}\pm \sqrt{\omega^2-p^2}\right)\tE(x,p)=0  \quad \mathrm{at} \quad x=\pm a \label{BC_WE-2D}
\end{equation}
for the Fourier transform (FT) $\tE(x,p)$ of the field $\E(x,z)$ with the respect to $z$.
Equation~(\ref{WE-2D}) is then Fourier transformed in the same manner, yielding
\begin{equation}
\left(\frac{d^2}{d x^2}+\epsilon\,\omega^2-p^2\right)\widetilde{\E}(x,p)=
-\omega^2 \widetilde{V}(x,p)\ast\widetilde{\E}(x,p)\,,
\label{WE-2D-FT}
\end{equation}
where $\widetilde{V}(x,p)$ is the FT of $V(x,z)=\varepsilon(x,z) - \epsilon$, the perturbation of the permittivity inside the BWG region, and $\ast$ denotes the convolution over $p$. We solve Eqs.\,(\ref{BC_WE-2D}) and (\ref{WE-2D-FT}) using the Green's function (GF) $G$ of the BWG for $|x|\leqslant a$, satisfying the equation
\begin{equation}
\left(\frac{d^2}{d x^2}+\epsilon\,\omega^2-\xi\right)G(x,x';\xi)=\delta(x-x')\label{VE_Gp2}
\end{equation}
and the BCs~\Eq{BC_WE-2D} at $x=\pm a$, where we have defined $\xi=p^2$.
This yields the integral equation
\begin{equation}
\widetilde{\E}(x,p)= -\omega^2 \int\limits_{-a}^a dx' G(x,x';p^2) \left(\widetilde{V}(x',p)\ast \widetilde{\E}(x',p)\right)\,.
\label{Epz}
\end{equation}
\begin{figure}
\includegraphics[width=\linewidth]{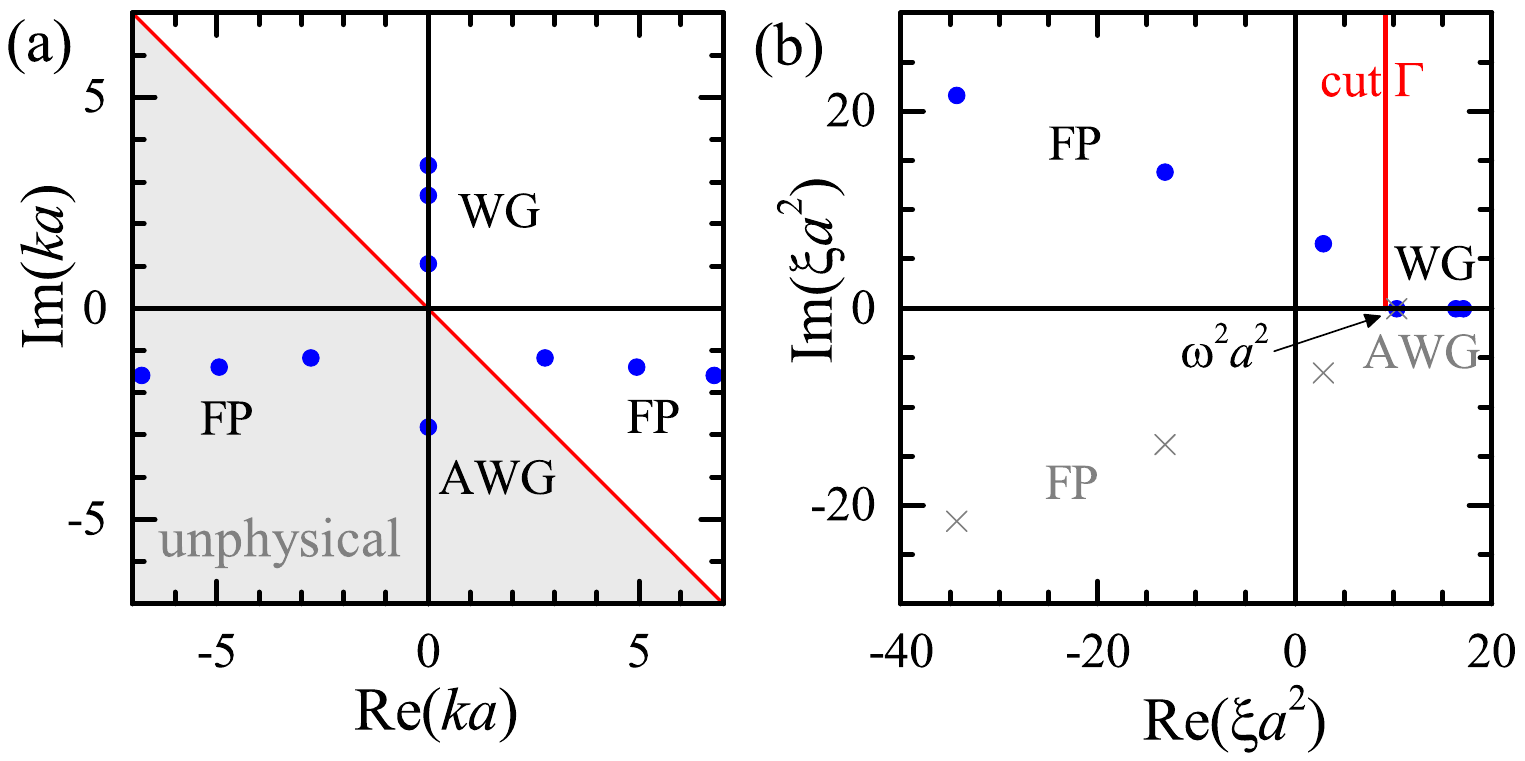}
\caption{(a) RS wave numbers -- poles of the GF of the BWG in the complex $k$-plane. The red line splits the $k$-plane into physical and unphysical half-planes, according to the cut $\Gamma$ in the $\xi$-plane.
(b)  Poles on the physical (circles) and unphysical (crosses) Riemann sheet  and the cut (red line) of the GF in the complex $\xi$-plane,
calculated for a photon energy $\hbar\omega$ of 3\,eV ($\omega a \approx 3.04$).
}\label{fig:ComplexPlane}
\end{figure}
Being considered in the complex $\xi$-plane,  $G(x,x';\xi)$ has {\it simple poles} at $\xi_n=\omega^2-k_n^2$, corresponding to RSs of the BWG, and, owing to the square root in the BCs~\Eq{BC_WE-2D}, a {\it cut} $\Gamma$, going from $\xi=\omega^2$ to infinity, and splitting the $\xi$-plane into two Riemann sheets. The GF has to be single valued and thus it is defined using only one of the Riemann sheets. This  ``physical'' sheet should respect the before mentioned outgoing boundary condition that $k=\sqrt{\omega^2-\xi}$ is positive real or positive imaginary on the real half-axis $\xi>0$. This requires that $\Gamma$ does not cross the $\xi>0$ half-axis, since if $\Gamma$ would cross the half-axis at $\xi_\mathrm{c}>0$, the right and left limits of $k(\xi)$ towards $\xi_\mathrm{c}$ have opposite signs, such that the condition to be positive real or positive imaginary cannot be fulfilled simultaneously for both limits.

\Fig{fig:ComplexPlane} shows a resulting mapping of the complex $k$-plane onto the complex $\xi$-plane. The cut $\Gamma$ is chosen here as a vertical half-axis (red line in \Fig{fig:ComplexPlane}(b), corresponding to the red line in \Fig{fig:ComplexPlane}(a)) which divides the $k$-plane into two half-planes, one of them corresponding to the physical sheet. The $k$-plane contains all possible values $k_n$ of RSs of the BWG, which include WG ($ik_n<0$), anti-WG ($ik_n>0$) and FP ($\Re(k_n)\neq0$) modes~\cite{Armitage14}. For the chosen BWG they are roots of the secular equation
\begin{equation}
(q_n-k_n)e^{2iq_n a}=(-1)^n (q_n+k_n), \label{Sol_Eq}
\end{equation}
where $q_n = \sqrt{(\epsilon-1)\omega^2+k_n^2}$ and $n$ is integer, see \App{sec:RSBWG} for details. Only a subset $\mathbb{S}$ of the RSs with $k_n$ values located on the physical half plane contributes as poles to the GF $G(x,x';\xi)$. Using the properties of the GF and applying the residue theorem we obtain the spectral representation of the GF (see \App{sec:GFk} for derivation):
\bea
&&G(x,x';\xi) = \SI_n \frac{E_n(x)E_n(x')}{\xi_n-\xi}\label{Gp2_0}
\\
&&\equiv \sum_{n \in \mathbb{S}} \frac{E_n(x)E_n(x')}{\xi_n-\xi} +
\int\limits_{\Gamma}d\xi'\sum_{\nu=\pm}
\frac{E_\nu(x;\xi')E_\nu(x';\xi')}{\xi'-\xi}\,,
\nonumber
\eea
where
\begin{gather}
E_n(x) = \frac{1}{2i^n}\sqrt{\frac{k_n}{k_n a+i}}\bigl(e^{iq_n x}+(-1)^n e^{-iq_n x}\bigl)\,, \label{E_n} \\
E_\pm(x;\xi)= \sqrt{\frac{k}{4\pi [\alpha^2\cos(2qa)\mp(q^2+k^2)]}} \left( e^{iqx}\pm e^{-iqx} \right), \label{E_pm}
\end{gather}
$\alpha = \omega\sqrt{\epsilon-1}$,
$k=\sqrt{\omega^2-\xi}$, $q = \sqrt{\epsilon\,\omega^2-\xi}$, and the integration is performed along the cut $\Gamma$, from the branch point $\xi=\omega^2$ to infinity ($\xi=\omega^2+i\infty$).

Equation~(\ref{Gp2_0}) determines a complete set (see \App{sec:GFk}) of basis functions inside the BWG, which consists of all the RSs on the physical sheet and a continuum of cut states. Here, the cut continuum is the remainder of the radiation continuum not taken into account by the FP modes on the physical sheet. Expanding the electric field $\E(x,z)$ inside the region $|x|\leqslant a$,
\be
{\E}(x,z) = \lefteqn{\sum_n} \int \,{A}_n(z) E_n(x)\,,
\label{E_expansion}
\ee
and substituting it into~\Eq{Epz} along with the spectral representation~\Eq{Gp2_0}, we obtain
\be
\lefteqn{\sum_n} \int \left( \widetilde{A}_n(p) + \omega^2 \frac{1}{p_n^2-p^2}
\lefteqn{\sum_m} \int\,\widetilde{V}_{nm}(p)\ast \widetilde{A}_m(p) \right) E_n(x) = 0, \label{SumBrakets}
\ee
where $\widetilde{V}_{nm}(p)=\int_{-a}^a E_n(x) \widetilde{V}_{nm}(x,p)E_m(x) dx$, $p_n^2=\xi_n$, and $\widetilde{A}_n(p)$ is the FT of the expansion coefficient $A_n(z)$.
To satisfy~\Eq{SumBrakets}, it is sufficient to require that
\begin{equation}
p^2\widetilde{A}_n(p) = p_n^2\widetilde{A}_n(p) + \omega^2 \lefteqn{\sum_m} \int \,\widetilde{V}_{nm}(p)\ast \widetilde{A}_m(p)\,.
\end{equation}
The inverse FT of this equation yields the {\it key equation} of the WG-RSE method:
\begin{equation}
-\frac{d^2}{dz^2} A_n(z) = p_n^2A_n(z) + \omega^2 \lefteqn{\sum_m}\int\, V_{nm}(z) A_m(z)
\label{d2An}
\end{equation}
in which the matrix elements of the perturbation $V_{nm}(z)$ are functions of $z$ only, the coordinate along the non-uniform WG, and are defined by
\begin{equation}
V_{nm}(z) = \int\limits_{-a}^a E_n(x)\bigl(\varepsilon(x,z)-\epsilon\bigr) E_m(x) dx\,.
\label{MEs}
\end{equation}
Notably, \Eq{d2An} is expected to be applicable also to WGs with a two-dimensional cross-section, such as fibres, for which the perturbation in \Eq{MEs} has to be integrated over the BWG cross section, and $\epsilon$ and $p_n$ referring to a suited BWG, such as a fibre with circular cross-section which is analytically treatable.

The formalism of the WG-RSE is also applicable in its present form to WGs with frequency dispersive inhomogeneities. Indeed, since the light frequency $\omega$ is fixed, the perturbation  $\varepsilon(x,z)-\epsilon$ of the permittivity in \Eq{MEs} can be taken as frequency dependent and complex, as illustrated in the example in \Sec{Sec:gold}.

\section{Applications of the WG-RSE}

The main equation of the WG-RSE, \Eq{d2An}, is an ordinary second-order matrix differential equation for the vector $A_n(z)$ of the amplitudes of the field expansion into the basis functions, which can be integrated analytically or numerically. For numerical integration, one can use a highly accurate finite-difference scheme, such as a fourth-order linear multistep algorithm~\cite{Numerov24}, recently implemented for solving a one-dimensional matrix Schr\"odinger's-like equation~\cite{Wilkes16}.

The analytic integration of \Eq{d2An} is possible in homogeneous regions of the non-uniform WG, in which $V_{nm}$ do not depend on $z$. In this case $A_n(z)$ become superpositions of $e^{\pm i\varkappa z} c_n$, where $\varkappa$ and $c_n$ are respectively the eigenvalues and eigenvectors of the linear matrix problem
\begin{equation}
	\lefteqn{\sum_m} \int \,(p^2_n\delta_{nm} + \omega^2 V_{nm}) c_m= \varkappa^2 c_n\,,
	\label{w-RSE2}
\end{equation}
which is the matrix equation of the {\em fixed-frequency RSE} for homogeneous planar WGs. Its convergence is studied in \App{sec:RSEConv}. The expansion coefficients of the eigenvectors in the propagation follow from Maxwell's BCs and can be found using the scattering matrix (S-Matrix) method, as it is done in the present work, see \App{sec:SMatrix} for details.

For the examples presented in this work, the permittivity and consequently the functions $V_{nm}(z)$ have a step-like form, defining regions of constant cross-section.
Therefore the fixed-frequency RSE determines the propagation wavevectors $\varkappa$ and the corresponding eigenvectors $c_n$ in each homogeneous region, while the S-Matrix solves~\Eq{d2An} over the whole structure.

Since there is a freedom in choosing the cut $\Gamma$, we defined it in such a way that its contribution is about minimized. Considering the analytic form of the cut functions, \Eq{E_pm}, it is clear that their normalization constants have the quickest exponential decrease if the cut starts from the branch point perpendicular to the real $\xi$-axis. While the cut path can be further optimized, e.g. by keeping a distance to FP modes which cause large cut amplitudes, in the present work we choose it simply along the imaginary $\xi$-axis as shown in \Fig{fig:ComplexPlane}(b).  As a result, the continuum of radiation modes is replaced by the FP modes in $\mathbb{S}$, offering a {\it natural discretization}, while the cut contribution is minimized. The remaining total pole weight of the cut, if treated as a stretched pole, is $C=C^++C^-$, where
\begin{equation}
C^\pm = \int\limits_{\Gamma}\left| d\xi
\frac{k a+i}{\pi [\alpha^2\cos(2qa)\mp(q^2+k^2)]}
\right|, \label{Eq:Cpm}
\end{equation}
resulting in values of 1.51, 0.48, and 0.69 for energies of 1\,eV, 3\,eV, and 5\,eV, respectively, for the BWG used in this work, see \Sec{subsec:WGhole}.

Conversely, when choosing the cut along the real axis, going to $-\infty$, $ \mathbb{S}$ contains WG modes only,
whereas the cut weight $C$ [see \Eq{Eq:Cpm}] actually diverges logarithmically.
This case corresponds to using WG and radiation modes only. Then the expansion~\Eq{E_expansion} is valid in the entire space, both inside and outside the BWG, and \Eq{d2An} can be obtained by substituting \Eq{E_expansion} directly into the wave equation~(\ref{WE-2D}) and using the standard orthonormality of modes given by the Hermitian inner product. Taking furthermore the limit $\epsilon\to1$ removes the WG modes from the expansion~\Eq{E_expansion}, leaving only the harmonic functions $\exp(ikx)\pm \exp(-ikx)$ of the cut. This corresponds to the FMM.

\subsection{Waveguide with hole}
\label{subsec:WGhole}

We now illustrate the WG-RSE on an example of a planar dielectric WG with a hole of length $L=900$\,nm and width $W=130$\,nm, at a distance $D=160$\,nm from the edge of the WG, as shown in \Fig{fig:Spectra}(a). As BWG we take the homogeneous part of this WG, with $a=200$\,nm and $\epsilon=2.4$. For the numerical calculations, we use a finite basis with $N=N_\mathrm{WG}+N_\mathrm{FP}+N_\mathrm{cut}$ basis states, which includes WG, FP, and cut modes, respectively.
The subset of FP modes is chosen by truncating the full set of FP modes on the physical sheet to $|k_n|<k_{\rm max}$, with a suitably chosen cut-off $k_{\rm max}$ while the subset of cut modes is produced by a discretization of the cut, as detailed in \App{sec:CutDisc}.

To demonstrate the efficiency of the WG-RSE, we calculate the S-Matrix $\hat{S}$~\cite{Tikhodeev02} containing the matrix elements $S_{ij}$ giving the complex amplitudes of scattering from incoming WG mode $j$ to outgoing WG mode $i$. The examples used in this work have equal WG modes on both sides, such that we can enumerate them using $i,j=1,2,\ldots,2N_\mathrm{WG}$ with the lower (higher) half refering to the modes on the left (right) side of the structure, respectively. The S-matrix determines the power scattering matrix $P_{ij}=|S_{ij}|^2$. Since the structures considered in this work have a mirror symmetry plane at $z=0$, we can write the power scattering matrix $\hat{P}$ as a symmetric matrix
\begin{equation}
\hat{P}=\begin{pmatrix}
\hat{R} & \hat{T} \\
\hat{T} & \hat{R}\,,
\end{pmatrix}
\end{equation}
which contains transmission $T_{ij}$ and reflection $R_{ij}$ coefficients with $i,j=1,2,\ldots,N_\mathrm{WG}$ with the WG modes enumerated with decreasing $p_n$. The calculated transmission coefficients $T_{ij}$ are shown in \Fig{fig:Spectra}(c) versus photon energy $\hbar\omega$.  We can see that the asymmetric hole in this WG allows up to 25\% power conversion from the fundamental (even) WG mode to the first excited (odd) mode. The electric field for excitation with the fundamental mode is given in \Fig{fig:Spectra}(b), illustrating this conversion.

\begin{figure}
\includegraphics[width=\linewidth]{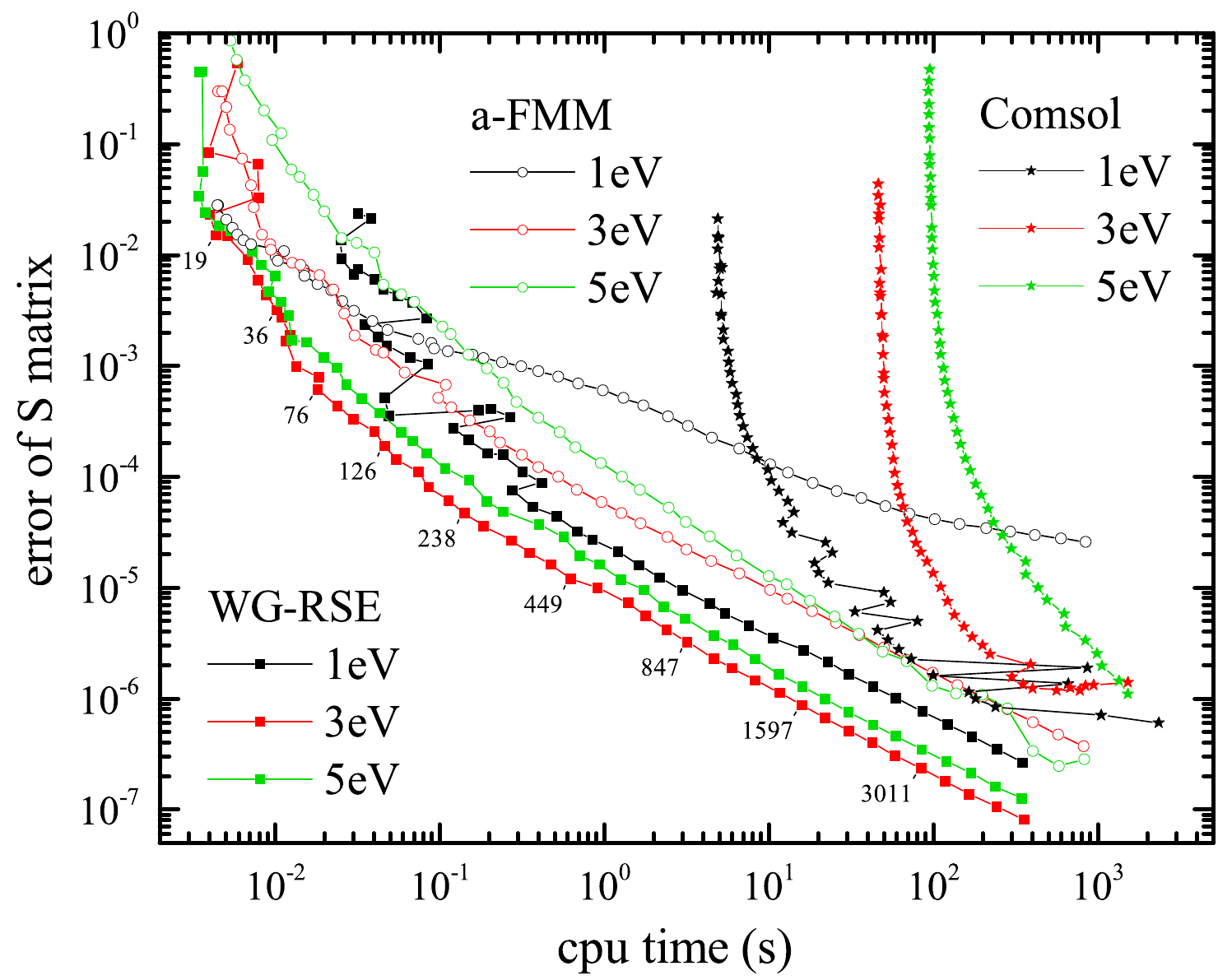}
\caption{Relative error of the S-matrix $\hat{S}$ versus computational time on a CPU Intel Core i7-5830K. Data is shown for WG-RSE, a-FMM and ComSol, and $\hbar\omega$ of 1, 3, and 5\,eV, as labelled. The basis size $N$ is indicated for the 3\,eV WG-RSE data.
}
\label{fig:Convergence}
\end{figure}

Since no analytical solution for $\hat{S}$ is available, we define the relative error\,%
\footnote{The norm of a matrix $\hat{A}$ is defined as $\Vert\hat{A}\Vert=\sup\limits_{\Vert{\bf x}\Vert=1}\Vert\hat{A}{\bf x}\Vert$ where $\Vert{\bf x}\Vert$ is the Euclidean norm of vector ${\bf x}$.} of the S-matrix as $\Vert\hat{S}-\hat{S}_0\Vert/\Vert\hat{S}_0\Vert$,
with respect to $\hat{S}_0$ calculated using the WG-RSE with the largest basis considered, $N=20000$.
\Fig{fig:Convergence} shows a comparison of the relative error of the WG-RSE with calculations using a-FMM and ComSol (see \App{sec:aFMMCOMSOL} for details).
We see that the WG-RSE is typically one to two orders of magnitude more efficient than ComSol and a-FMM. It is important to note that this conclusion does not depend on the choice to use $\hat{S}_0$ calculated by the WG-RSE. All methods eventually reach an error below $10^{-6}$, so that for errors $\gg 10^{-6}$ the results are independent of this choice. We show this explicitly later in \Fig{fig:ConvergenceAuComSol}.

\begin{figure}
	\includegraphics[width=\linewidth]{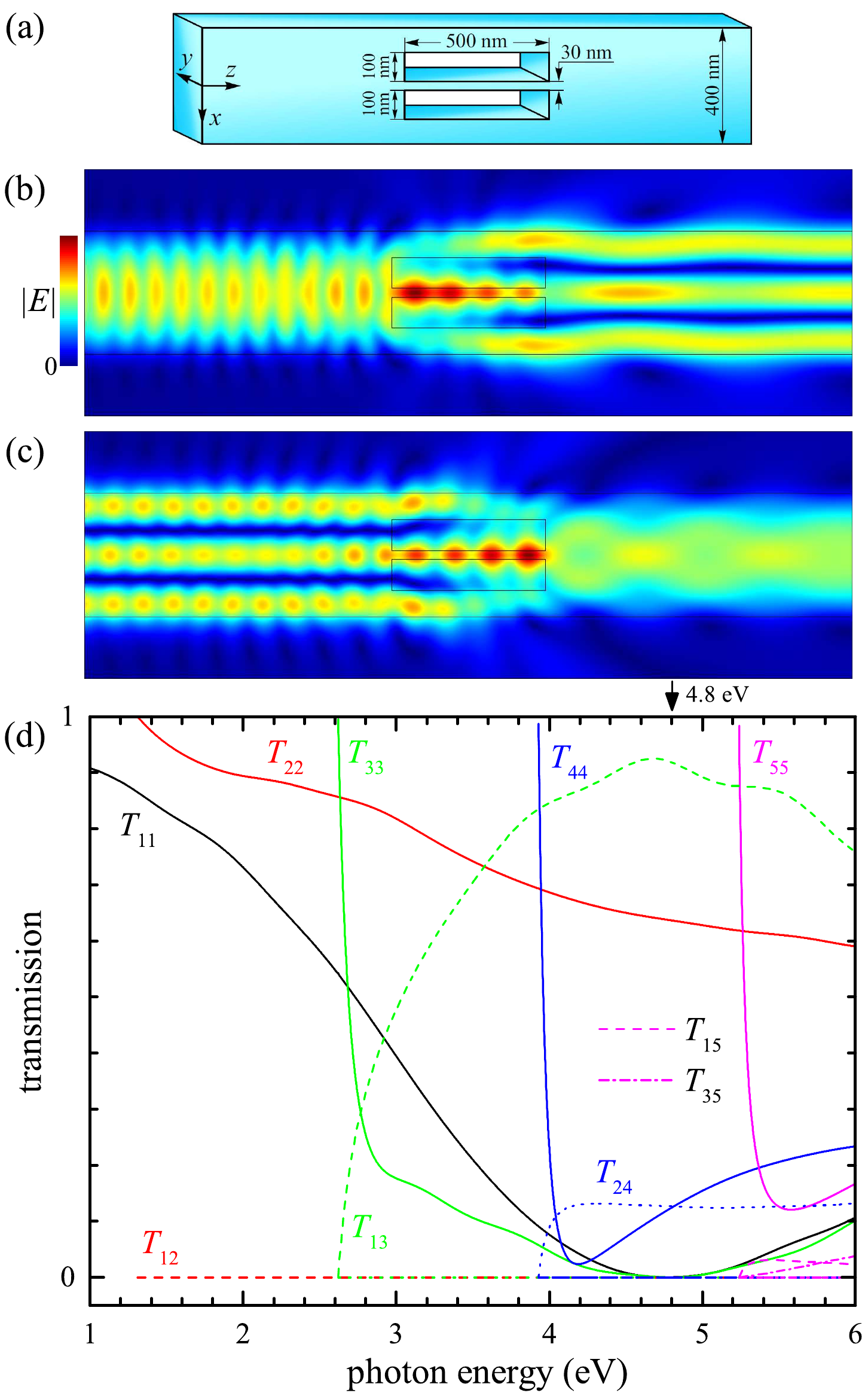}
	\caption{As \Fig{fig:Spectra} but for a double hole as sketched in (a). The structure has mirror symmetry with respect to the plane $x=0$. (b,c) Electric field amplitude for excitation with WG mode 1 (b) and 3 (c), at $\hbar\omega=4.783$\,eV from the left, on a linear color scale as given, overlaid with the WG outline. (d) Relative power transmission $T_{ij}$, from left WG mode $j$ to right WG mode $i$, as function of the photon energy.}
	\label{fig:SpectraDev5}
\end{figure}

\subsection{Waveguide with double hole}
\label{subsec:doublehole}
As a second example for the calculated transmission using the WG-RSE method, we show here the results for a double hole perturbation of a waveguide. The structure is shown in \Fig{fig:SpectraDev5}(a), having mirror symmetry about the $x=0$ plane. The resulting transmission in \Fig{fig:SpectraDev5}(d) shows selection rules as no conversion between WG modes of different parity is occurring, e.g. $T_{12}=0$. For a photon energy of 4.8\,eV, a nearly complete conversion between modes 1 to 3 is found (note that $T_{ij}=T_{ji}$ for systems with mirror symmetry about the $z=0$ plane). This is illustrated by the field distributions for excitation with WG mode 1 in \Fig{fig:SpectraDev5}(b) and WG mode 3 in \Fig{fig:SpectraDev5}(c).

\subsection{Waveguide with gold bar}
\label{Sec:gold}

\begin{figure}
	\includegraphics[width=\linewidth]{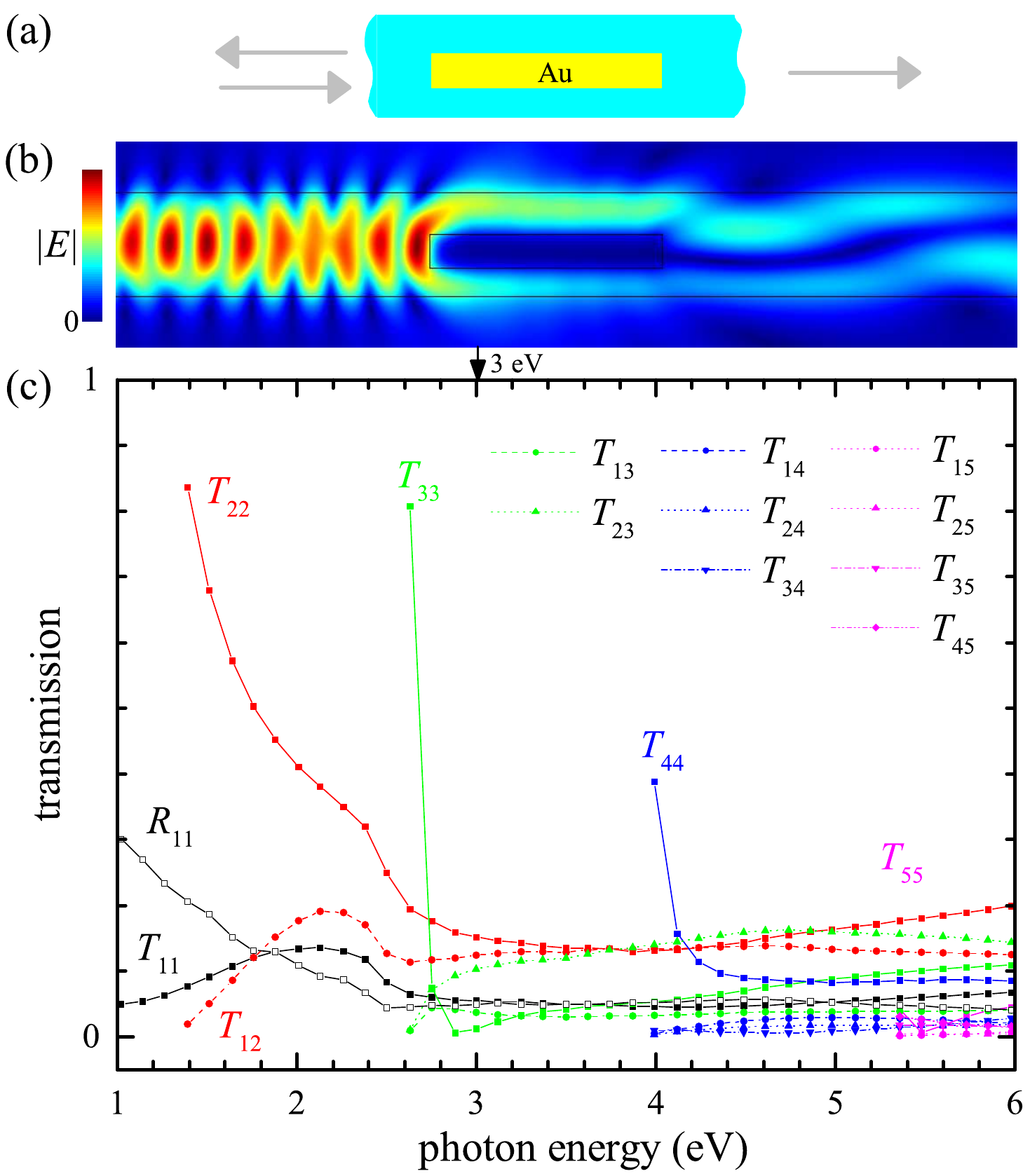}
	\caption{As \Fig{fig:Spectra}, but filling the hole in the waveguide with gold. The discrete energies used are the ones tabulated in \Onlinecite{JohnsonPRB72}.}
	\label{fig:SpectraAu}
\end{figure}

As example for a strongly dispersive and absorptive material as perturbation, we fill the hole in the waveguide of \Sec{subsec:WGhole} with gold. Since the WG-RSE uses a fixed frequency, the dispersion of the susceptibility is not relevant for the results. However, replacing vacuum with gold creates a very strong and absorptive perturbation. The resulting field distribution and transmission coefficients of the S-Matrix are shown in \Fig{fig:SpectraAu}. The data was calculated at the spectral points for which the susceptibility was measured in \Onlinecite{JohnsonPRB72}. We can see that the gold bar leads to a significant reflection, visible by the standing wave pattern in \Fig{fig:SpectraAu}(b) and in the reflection coeffcient $R_{11}$ shown in \Fig{fig:SpectraAu}(c). The transmission of the fundamental mode is accordingly low, in the 10\% range.  The second order mode instead has a higher transmission as it has a node in the region of the gold bar.

\begin{figure}
	\includegraphics[width=\linewidth]{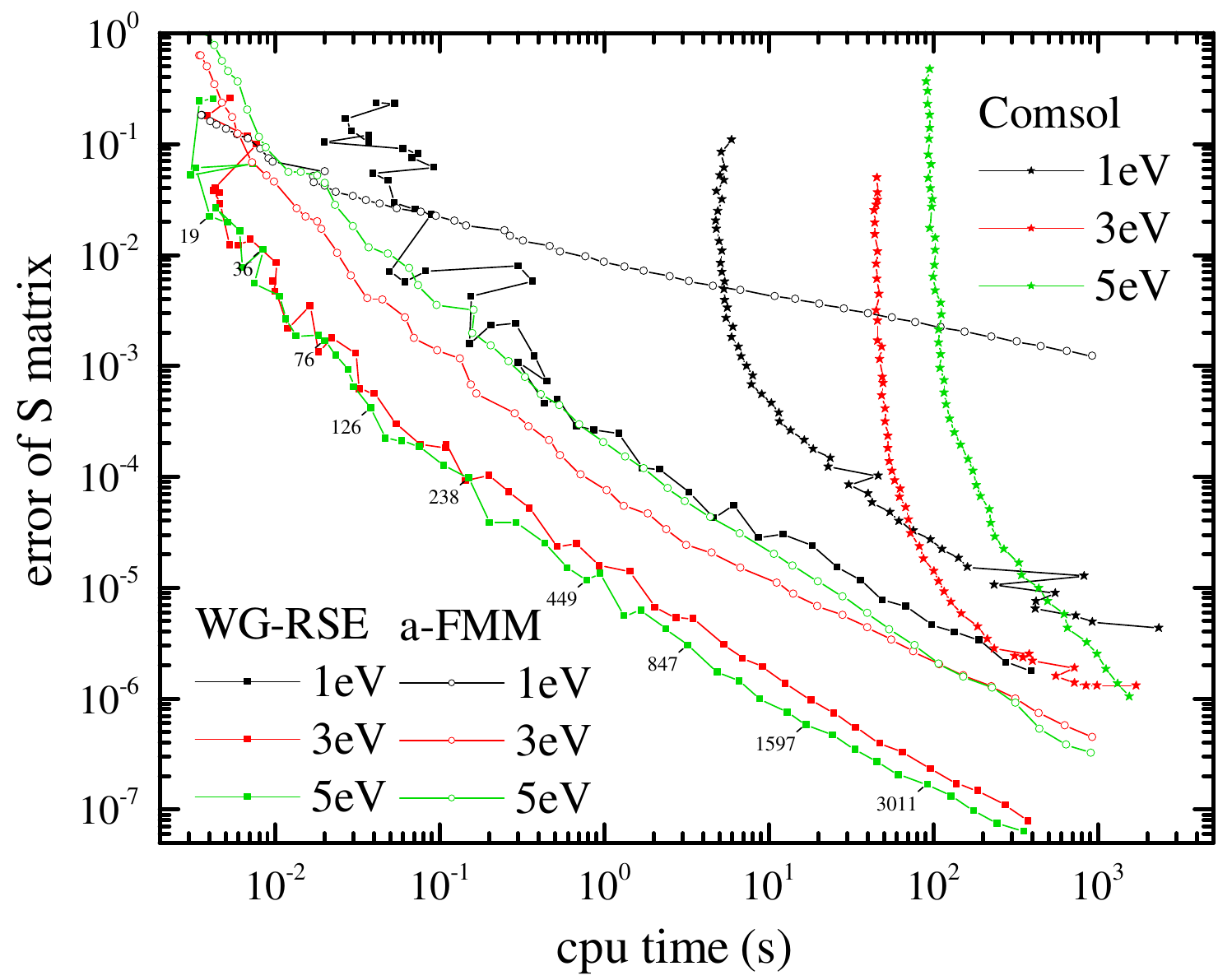}
	\caption{As \Fig{fig:Convergence}, but filling the hole in the waveguide with gold.}
	\label{fig:ConvergenceAu}
\end{figure}

\begin{figure}
	\includegraphics[width=\linewidth]{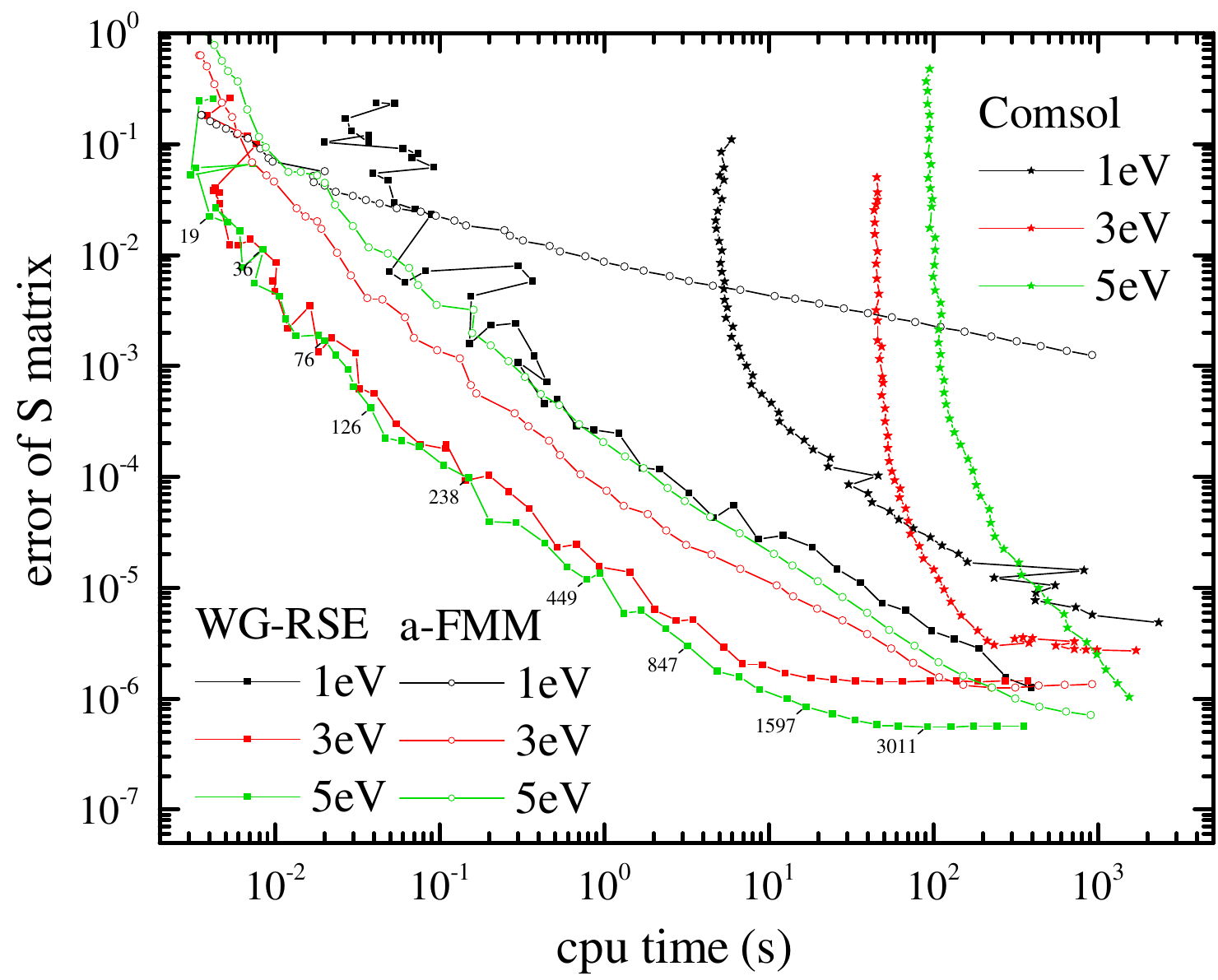}
	\caption{As \Fig{fig:ConvergenceAu}, but using as reference $\hat{S}_0$ the ComSol solutions with $m = 6$ (see \App{sec:aFMMCOMSOL}).}
	\label{fig:ConvergenceAuComSol}
\end{figure}

The corresponding convergence is shown in \Fig{fig:ConvergenceAu} and \Fig{fig:ConvergenceAuComSol}, using as $\hat{S}_0$ the highest accuracy WG-RSE or ComSol calculation, respectively. both display similar features as the air hole example \Fig{fig:Convergence}. Again, we find that the WG-RSE has a 1-2 orders of magnitude higher numerical efficiency.

\subsection{Waveguide with resonant cavity}

\begin{figure}
	\includegraphics[width=\linewidth]{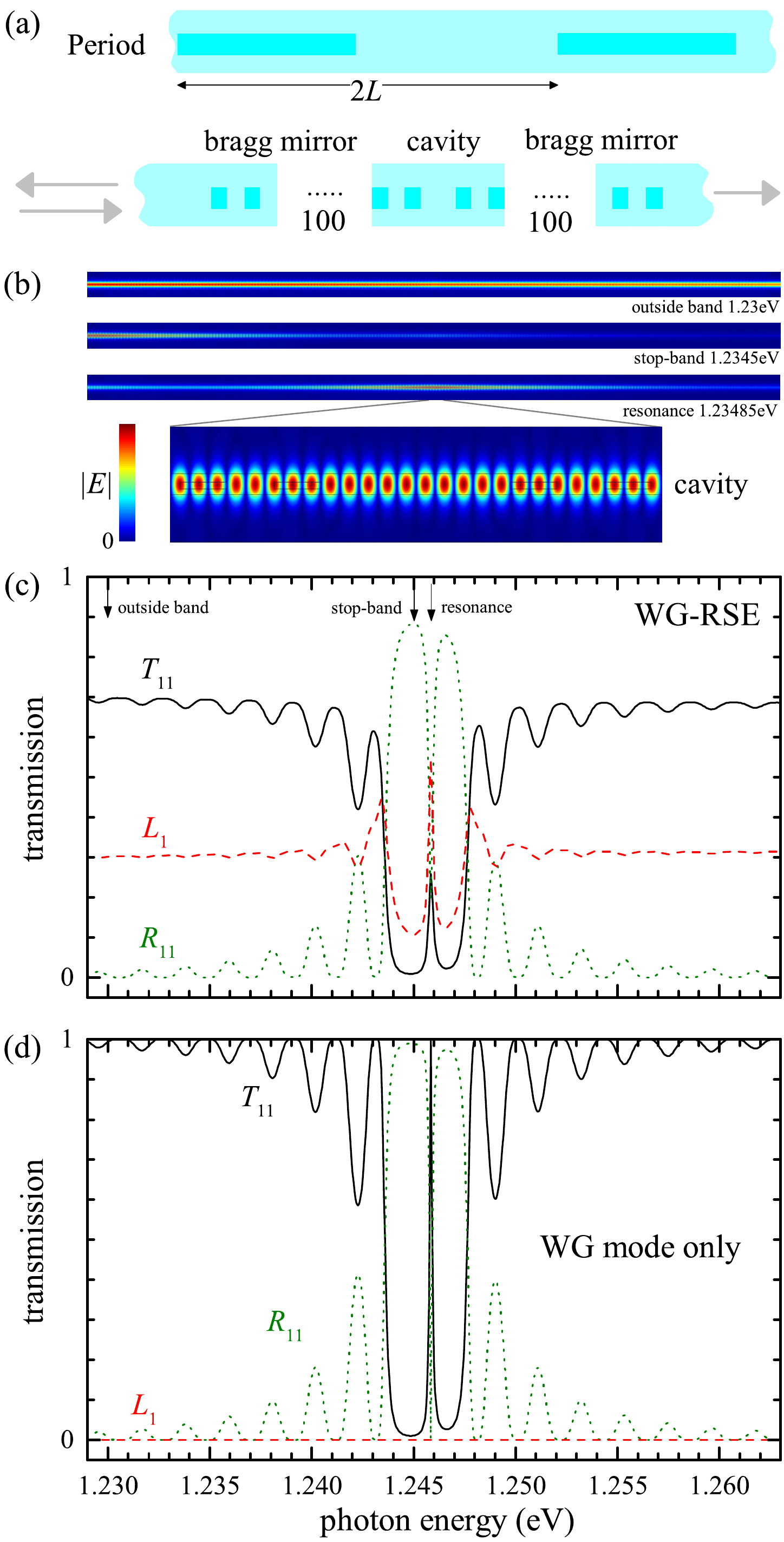}
	\caption{As \Fig{fig:Spectra}, but filling the hole in the waveguide with $\epsilon=2.6$, and creating a  cavity structure made of two 100 period Bragg mirrors of this perturbation with a period of $2L=1800$\,nm, surrounding a cavity of length $2L$. (a) Schematic of the structure. (b) Electric field amplitude for excitation with WG mode 1 propagating from the left, on a linear color scale as given, overlaid with the WG outline. Data outside the stop-band at $\hbar\omega=1.23$\,eV, in the stop-band at $\hbar\omega=1.245$\,eV, and at the cavity resonance $\hbar\omega=1.24585$\,eV. (c) Relative power transmission $T_{11}$, reflection $R_{11}$, and loss $L_1=1-T_{11}-R_{11}$, as function of the photon energy $\hbar\omega$. (d) as (c), but restricting the S-Matrix calculation to the WG mode.}
	\label{fig:SpectraMC}
\end{figure}

As an example of an extended non-uniform WG, we chose a cavity structure with two Bragg mirrors of 100 periods each, shown in \Fig{fig:SpectraMC}(a). Each period consists of the hole in the waveguide of \Sec{subsec:WGhole} filled with a material of $\epsilon=2.6$, close to the $\epsilon=2.4$ of the waveguide, followed by a waveguide section of equal length $L$. The cavity is formed by a waveguide section of length $2L$, surrounded by the Bragg mirrors. Choosing such a small perturbation reduces the scattering losses. This structure has a cavity resonance at $1.24585$\,eV, for which the waveguide is single-moded, i.e. it supports only one waveguide mode.
The calculated electric field for three photon energies is shown in (b). Outside the Bragg stop band ($\hbar\omega=1.23$\,eV), the field is rather homogeneous, inside the Bragg stop band ($\hbar\omega=1.245$\,eV) the field is decaying as it gets reflected, and at resonance with the cavity mode at $\hbar\omega=1.24585$\,eV a resonant enhancement in the cavity is observed. The calculated transmission $T_{11}$, reflection $R_{11}$, and losses $L_1=1-T_{11}-R_{11}$, are given in (c) for the WG-RSE using $N=2000$. The Bragg stop band of about 5\,meV width is evident, hosting a resonance at $1.24585$\,eV with a Q-factor of about 6000.

The loss $L_1$ is significant, about 30\%, reducing to 11\% in the stop band and increasing to 54\% at resonance. The loss results from scattering into non-WG modes by the large number of interfaces present. The observed loss reduction in the stop-band is due to the lower penetration of the light into the structure, and the enhancement at resonance is due to the increased field inside the structure.

\begin{figure}
	\includegraphics[width=\linewidth]{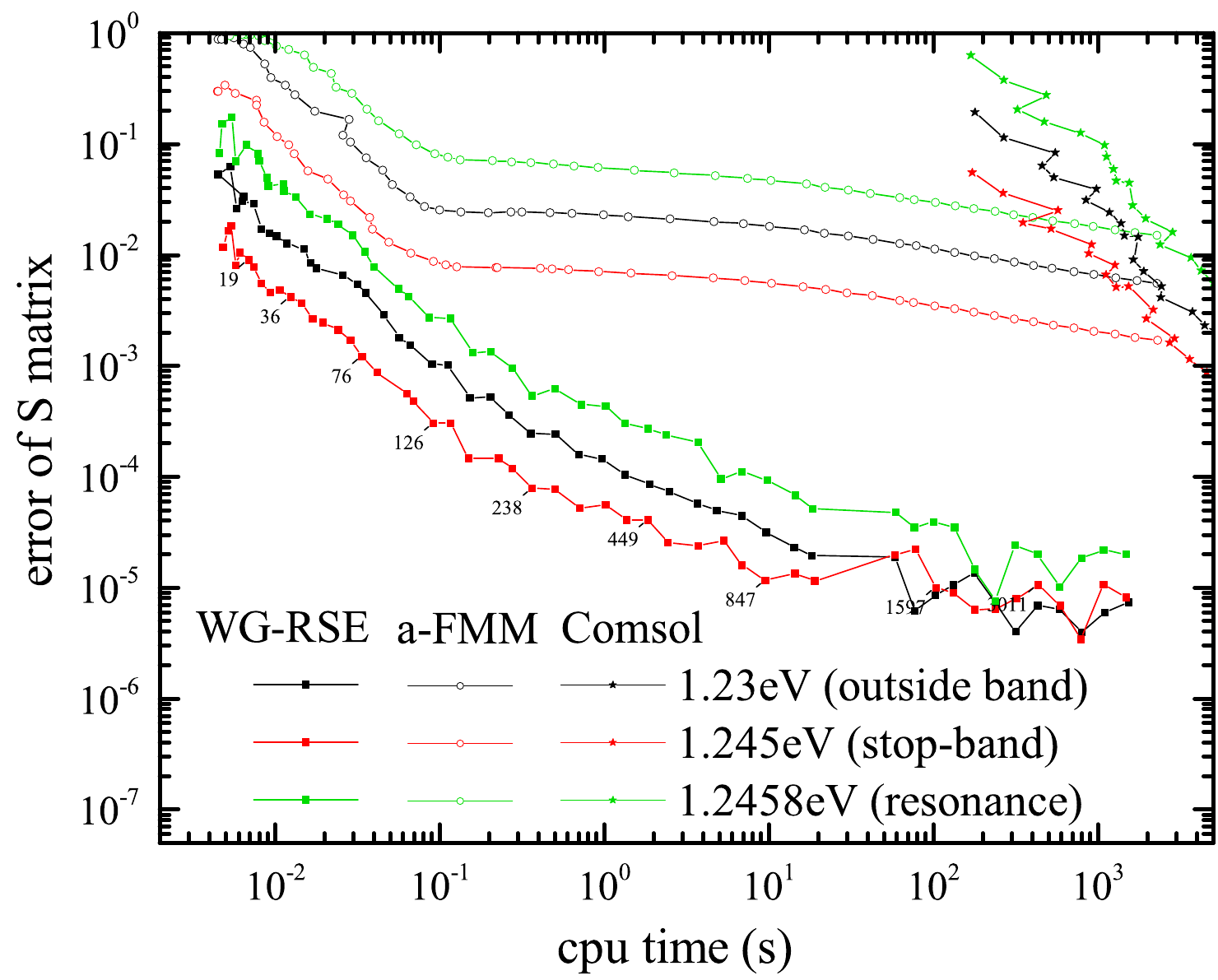}
	\caption{As \Fig{fig:Convergence}, but for the waveguide cavity structure and the three photon energies outside the stop-band at 1.23\,eV, in the stop-band at 1.245\,eV, and on resonance at 1.24585\,eV, see \Fig{fig:SpectraMC}.
The reference $2\times2$ scattering matrix $\hat{S}_0$ was calculated using the WG-RSE with the largest basis considered, $N=20000$.
}
\label{fig:ConvergenceMC}
\end{figure}

We emphasize that the total length of this structure is about $360\,\mu$m, corresponding to 363 free-space resonant wavelengths, a large simulation space for FEM solvers, making them inefficient. This is exemplified in the comparison of the errors of the WG-RSE, a-FMM and ComSol in \Fig{fig:ConvergenceMC}. We observe that the ComSol calculation times are around 5 orders of magnitude longer than the WG-RSE for equal errors. This time is dominated by the time to create the calculation grid, but even with the grid already build (not shown) the computation is still about 4 orders of magnitude longer than the WG-RSE. This exemplifies the advantage of the WG-RSE in calculating such extended structures containing fine detail. For the a-FMM we find for large errors a similar result as before, being about 1 order of magnitude slower than the WG-RSE. For small errors however, the a-FMM convergence slows down dramatically. We note that these results were obtained using the same PML distance and width settings as function of basis size as for the previous examples. It is possible to optimize, specific to each energy, the PML parameters to provide a faster convergence, which in some cases can become comparable to the WG-RSE. However, such an approach is computationally inefficient as a dependence on the PML parameters needs to be explored in each case, and the convergence behaviour is not uniform. For example, relying only on the a-FMM results in the present example, one could be mislead to the conclusion that the results were converged at about 0.1\,s cpu time, since the a-FMM results remain effectively constant over the next two orders of cpu time. However, at this point the actual errors are still amounting to a few percent.

A simple approach used in the literature to treat such long structures is reducing the calculation only to the WG modes of the constant sections. To show the result of such a treatment, we have limited the S-Matrix calculation to the only bound mode which exists in each part of the structure for the frequency range considered. Specifically, after solving eigenvalue problem~\Eq{w-RSE2}, we replace the expansion~\Eq{E_expansion} by only one element -- the WG mode of the BWG, and leave only WG modes in the scattering matrix $\hat{S}$ \Eq{Eq:Smatrix_b}.
The result is shown in \Fig{fig:SpectraMC}(d). While the cavity resonance and the stop band width is reproduced well, the losses are not treated correctly -- they are not present in this model. Accordingly, we find $L_1=0$ and a sharper cavity resonance, with a Q-factor of about 9000. This is expected, as the missing non-WG modes are disabling the losses, so that the resonance width is solely determined by the Bragg mirror reflectivities.

\section{Conclusions}

In conclusion, we have developed a waveguide resonant-state expansion (WG-RSE), a new general method, based on the concept of resonant states, for calculating light propagation in waveguides with varying cross-section. We have shown the fundamental importance of resonant states which provide a natural discretization of the continuum of light waves scattered by the waveguide inhomogeneities, thus building an optimal basis for expansion of the electromagnetic field. As a result, the WG-RSE can be orders of magnitude more computationally efficient than present state of the art methods, such as the aperiodic Fourier modal or finite element method, as we have demonstrated on several examples of non-uniform planar waveguides. Importantly, the WG-RSE is transferable to other fields of physics showing wave phenomena, such as acoustics and quantum mechanics, enabling a wide application perspective.

\begin{acknowledgments}
This work was supported by the Cardiff University EPSRC Impact Acceleration Account EP/K503988/1 and the S\^er Cymru National Research Network in Advanced Engineering and Materials.
\end{acknowledgments}

\appendix

\section{Resonant states of the basis waveguide}
\label{sec:RSBWG}

Resonant states (RSs) of the basis waveguide (BWG) are solutions of the wave equation
\be
\left(\frac{d^2}{dx^2}+\alpha^2+k_n^2\right)E_n(x)=0 \quad \mathrm{for} \quad |x|<a\,
\label{VolEq_En}
\ee
with outgoing or incoming BCs
\be	\left(i\frac{d}{dx}\pm k_n\right)E_n(x)=0  \quad \mathrm{for} \quad x=\pm a\,,\ee
where $\alpha = \omega\sqrt{\epsilon-1}$ and $n$ is the integer index which labels the RSs. The electric field of the $n$-th RS has the form
\begin{equation}
	E_n(x) = C_n\bigl(e^{iq_n x}+(-1)^n e^{-iq_n x}\bigr)\,,
	\label{E_nX}
\end{equation}
where $q_n = \sqrt{\alpha^2+k_n^2}$, and the RS wave numbers $k_n$ are determined by the secular equation \Eq{Sol_Eq}, following from Mawxell's BCs.
Note that the solutions of \Eq{Sol_Eq} with ${k_n=\pm i \alpha}$, $q_n=0$, and odd $n$ should be excluded from the set of the eigenvalues since they corresponds to zero electric field. The normalization constants $C_n$ are given by
\begin{equation}
	C_n = \frac{1}{2i^n}\sqrt{\frac{k_n}{k_n a+i}}
\end{equation}
and are found from the orthonormality of RSs, which for the fixed frequency problem treated here is given by
\begin{multline}
	\int\limits_{-a}^a E_n(x)E_m(x)dx\,\\
	-\,\frac{E_n(a)E_m(a)+E_n(-a)E_m(-a)}{i(k_n+k_m)}=\delta_{nm}\,,
\label{norm}
\end{multline}
where $\delta_{nm}$ is the Kronecker symbol. Note that \Eq{norm} is obtained following the general procedure for normalizing RSs as outlined in \cite{Muljarov10,Doost14}. Interestingly, the normalization condition \Eq{norm} for RSs at a fixed frequency does not contain in the volume term the permittivity of the system $\varepsilon(x)$ as a weight function, unlike RSs of a planar systems defined at a fixed in-plane wave vector~\cite{Muljarov10,Armitage14}.

\section{Green's function of the basis waveguide}
\label{sec:GFk}

\begin{figure}
	\includegraphics[width=\linewidth]{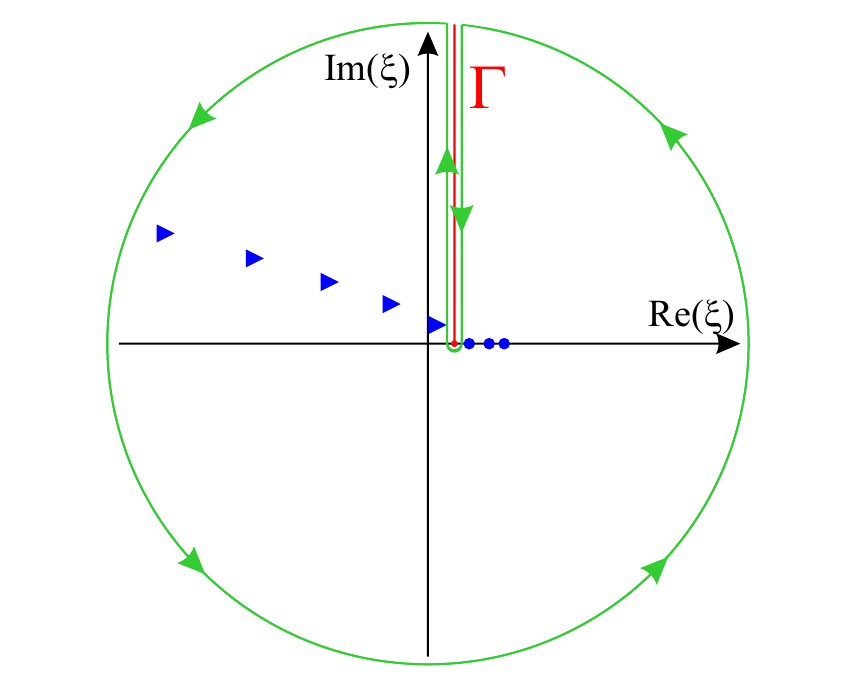}
	\caption{``Physical'' Riemann $\xi$-sheet.
		Symbols mark poles of the Green's function $G(x,x';\xi)$ of the BWG.
		The cut $\Gamma$ starting from the branch point at $\xi=\omega^2$ is shown by a red line.
		Green curve marks the path of integration.}\label{fig:CP_intPath}
\end{figure}

The Green's function (GF) $G(x,x';\xi)$ of the BWG satisfies the wave equation
\be \left(\frac{d^2}{d x^2}+\epsilon\,\omega^2-\xi\right)G(x,x';\xi)=\delta(x-x') \quad \mathrm{for} \quad |x|<a\ee
and the BCs
\be \left(i\frac{d}{d x}\pm \sqrt{\omega^2-\xi}\right)G(x,x';\xi)=0  \quad \mathrm{for} \quad x=\pm a  \,. \label{BC_WE-2DX} \ee
It has the analytic form
\begin{equation}
	G(x,x';\xi) = - \frac{e_L(x_<,k)e_R(x_>,k)}{W(k)}, \label{Greens_function}
\end{equation}
where $x_<=\min(x,x')$, $x_>=\max(x,x')$, and $e_L(x,k)$ and $e_R(x,k)$ are solutions of the corresponding homogeneous wave equation satisfying, respectively, the left (at $x=-a$) and the right (at $x=a$) BC. For the constant permittivity $\epsilon$ of the slab these solutions have the explicit analytic form
\be
e_{L,R} (x,k) = \pm \frac{e_+(x)}{N_+(k)} + \frac{e_-(x)}{N_-(k)} \,,
\label{Eb}
\ee
where
\bea
e_\pm(x,q) &= &e^{iqx} \pm e^{-iqx}\,,
\label{Epm}
\\
N_\pm(k) &=& (q-k)e^{iqa}\mp (q+k)e^{-iqa}\,,\\
k&=&\sqrt{\omega^2-\xi}\,,
\label{kdef}
\\
q&=&\sqrt{\alpha^2+k^2}\,,
\label{qdef}
\eea
and the Wronskian $W(k)$ is given by
\be
W(k) = e'_L(x) e_R(x) - e_L(x) e'_R(x)=\frac{-8iq}{N_+(k) N_-(k)}\,.
\label{Wronskian}
\ee
Note that the GF is invariant with respect to the sign of $q$, but changes its value if the sign of $k$ in \Eq{Greens_function} changes to the opposite. Therefore the square root which appears in the definition of $k$, originating from the BCs~\Eq{BC_WE-2D}, produces a cut of the GF in the complex $\xi$ plane, going from the branch point at $\xi=\omega^2$ (corresponding to $k=0$) to infinity.  The difference in the values of the GF on different sides of the cut is then given by
\bea
\Delta G(x,x',\xi)&=&-\frac{e_L(x_<,k)e_R(x_>,k)}{W(k)}\nonumber\\
&&+\frac{e_L(x_<,-k)e_R(x_>,-k)}{W(-k)}
\label{cut} \\
&=& -2\pi i\sum_{\nu=\pm}\sigma_\nu(\xi) e_\nu(x_<,q) e_\nu(x_>,q)\,,
\nonumber
\eea
where
\be
\sigma_\pm(\xi)=\frac{k}{4\pi [\alpha^2\cos(2qa)\mp(q^2+k^2)]}\,.
\label{sigma}
\ee
In addition to the cut, the GF has simple poles in the complex $\xi$ plane, at $\xi=\xi_n$, determined by the equation $W(k)=0$, equivalent to~\Eq{Sol_Eq}. The residues of the GF~\Eq{Greens_function} at these poles are
\be
{\rm Res}\, G(x,x';\xi_n)= -\frac{(-1)^n k_n}{4(k_na+i)} e_\pm(x,q_n)e_\pm(x',q_n)\,,
\label{Residue}
\ee
using $e_+(x,q_n)$ for even $n$ and $e_-(x,q_n)$ for odd $n$.
Now, choosing the cut $\Gamma$  as shown in Fig.\,\ref{fig:CP_intPath} and selecting the ``physical'' Riemann sheet, which contains the waveguide (WG) modes and the Fabry-Perot (FP) modes with Re$(k)>0$,  we apply the residue theorem to the function $G(x,x';\xi')/(\xi-\xi')$, integrating it in the complex $\xi'$ plane, along a closed contour shown in \Fig{fig:CP_intPath}. The contour consists of a large circle, two parallel lines circumventing the cut, and a vanishing half-circle surrounding the branch point. Since the GF is vanishing at large $\xi'$ and is finite at the branch point, both circular integrals vanish, and the residue theorem yields
\be
G(x,x';\xi)=\sum_{n\in \mathbb{S}} \frac{{\rm Res}\, G(x,x';\xi_n)}{\xi-\xi_n}\, -\!\!\!\!\!\int\limits_{\omega^2}^{\omega^2+i\infty} \!\!\!\frac{\Delta G(x,x',\xi')d\xi'}{2\pi i(\xi-\xi')}.
\label{GF-residue}
\ee
Here $\mathbb{S}$ includes all the poles of the GF inside the closed contour, i.e. all poles on the selected physical sheet, and the integration in the second term is performed along the cut. Using Eqs.\,(\ref{cut}) and (\ref{Residue}), this results in Eqs.\,(\ref{Gp2_0})--(\ref{E_pm}).

Note that the discrete modes $E_n(x)$ and the cut modes $E_\pm(x;\xi)$ together constitute a complete set of basis functions, suitable for expansion of an arbitrary field within the region $|x|\leqslant a$.
This can be seen by substituting the series \Eq{Gp2_0} into \Eq{VE_Gp2} and using \Eq{VolEq_En}, valid for both discrete and cut modes, which yields the closure relation

\be
\lefteqn{\sum_n} \int E_n(x)E_n(x')=\delta(x-x')\,.
\label{closure}
\ee

\section{Discretization of the cut}\label{sec:Discretization}
\label{sec:CutDisc}
\begin{figure}[b]
	\includegraphics[width=\linewidth]{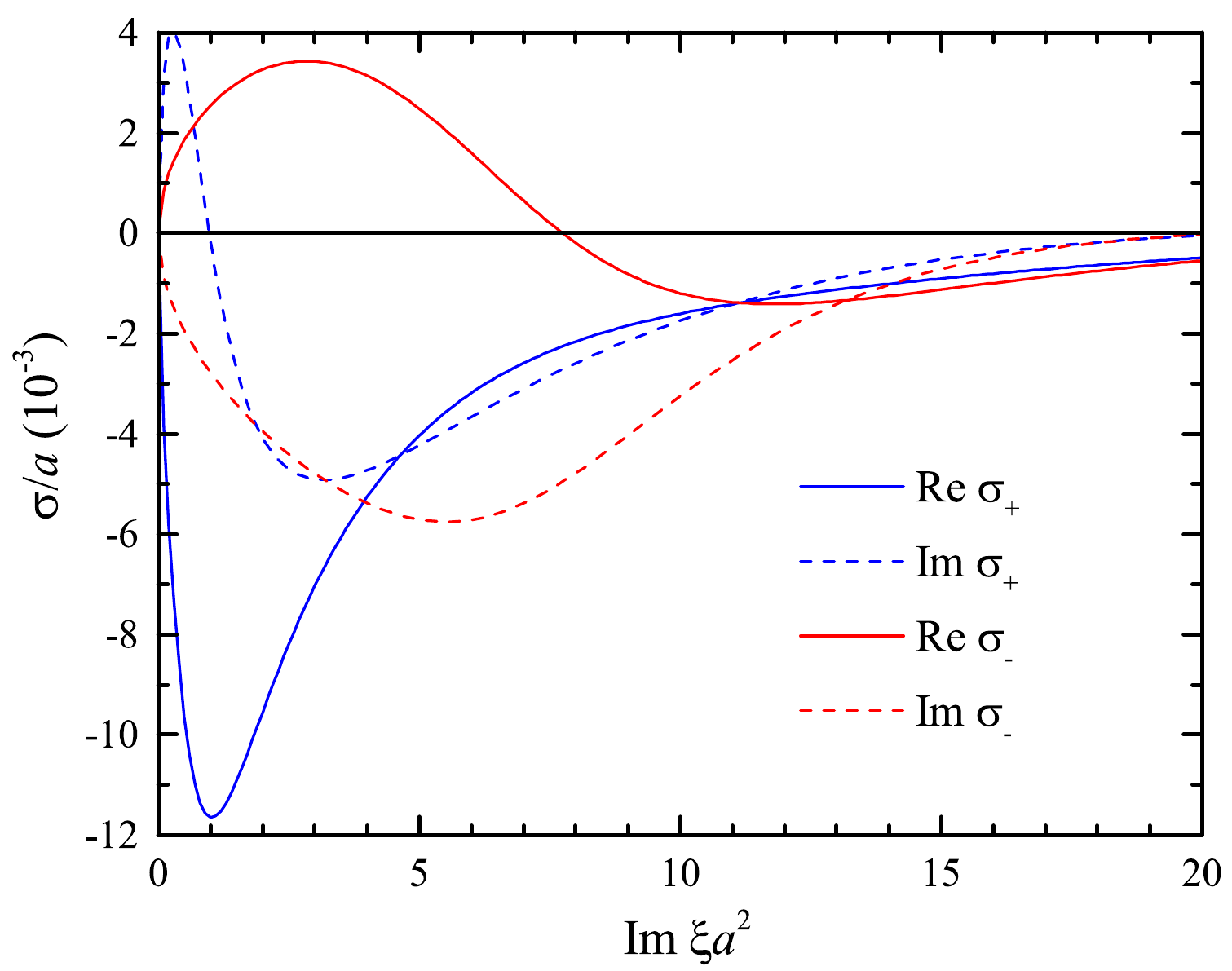}
	\caption{Cut weight $\sigma_\pm$ along the cut $\Gamma$ for a photon energy of $\hbar\omega=3$\,eV. Other parameters as in Fig.\,1 of the main text}\label{fig:CutWeight}
\end{figure}

In numerical calculations, we discretize the remainder of the continuum of radiation modes, represented by the cut $\Gamma$ in \Fig{fig:CP_intPath} (an example of the cut weight $\sigma_\pm$ is shown in \Fig{fig:CutWeight}), by replacing the cut with a finite number of  poles which we add to the basis of RSs along with the normal RSs included in $\mathbb{S}$. This is done following a similar procedure as described in \Onlinecite{Doost13}.  Namely, we first split the cut modes into even and odd subgroups, labeled by $\nu=+$ and $\nu=-$, respectively. Then, for each subgroup, we divide the cut into $N_\mathrm{cut}^\nu$ intervals $[\xi^\nu_n,\xi^\nu_{n+1}]$ with an equal weight defined as
\begin{equation}
	w_\nu=\int\limits_{\xi^\nu_n}^{\xi^\nu_{n+1}}|\sqrt{\sigma_\nu(\xi)}|\,d\xi\,,
\end{equation}
in this way determining the points $\xi^\nu_n$,  where $n=1,2,\ldots N_\mathrm{cut}^\nu$, $N_\mathrm{cut}=N_\mathrm{cut}^++N_\mathrm{cut}^-$, $\xi^\nu_1=\omega^2$, and $\xi^\nu_{N_\mathrm{cut}^\nu+1}=\omega^2+i\infty$.
Note that the normalization constants of the cut states~\Eq{E_pm} are given by $\sqrt{\sigma_\nu(\xi)}$.
Each interval $[\xi^\nu_n,\xi^\nu_{n+1}]$ is then replaced by a fictitious RS having the wave function given, as in~\Eq{E_n}, by
\begin{equation}
	\tilde{E}^\nu_n(x) = \tilde{C}^\nu_n\bigl(e^{i\tilde{q}_n x}+\nu e^{-i\tilde{q}_n x}\bigr)\,,
	\label{tildeE_n}
\end{equation}
where the coefficients $\tilde{C}_n^\nu$  and the positions $\tilde{\xi}_n^\nu$ of the fictitious poles are defined by
\begin{gather}
	\tilde{C}_n^\nu = \left(\int\limits_{\xi^\nu_n}^{\xi^\nu_{n+1}} \sigma_\nu(\xi)\,d\xi\right)^{1/2},\\
	\tilde{\xi}_n^\nu = \frac{1}{(\tilde{C}_n^\nu)^2}\int\limits_{\xi^\nu_n}^{\xi^\nu_{n+1}}
	\sigma_\nu(\xi)\xi \,d\xi\,,
\end{gather}
and $\tilde{q}_n$ and $\tilde{k}_n$ are given by Eqs.\,(\ref{kdef}) and (\ref{qdef}). The resulting fictitious RSs produce a set of modes $\tilde{\mathbb{S}}$ which we add to the basis of RSs $\mathbb{S}$ and treat the resulting discrete matrix problem numerically.
The final basis consists of $N=N_\mathrm{WG}+N_\mathrm{FP}+N_\mathrm{cut}$ basis states which include WG, FP, and cut modes, respectively.
In numerical calculations, we use a ratio between FP and cut poles of $N_\mathrm{FP}/N_\mathrm{cut} \approx (\omega a)/(2 \log N)$, which we found to approximately minimize the errors for a given basis size $N$.

\section{Convergence of the fixed frequency RSE}
\label{sec:RSEConv}

\begin{figure}[htb]
	\includegraphics[width=\linewidth]{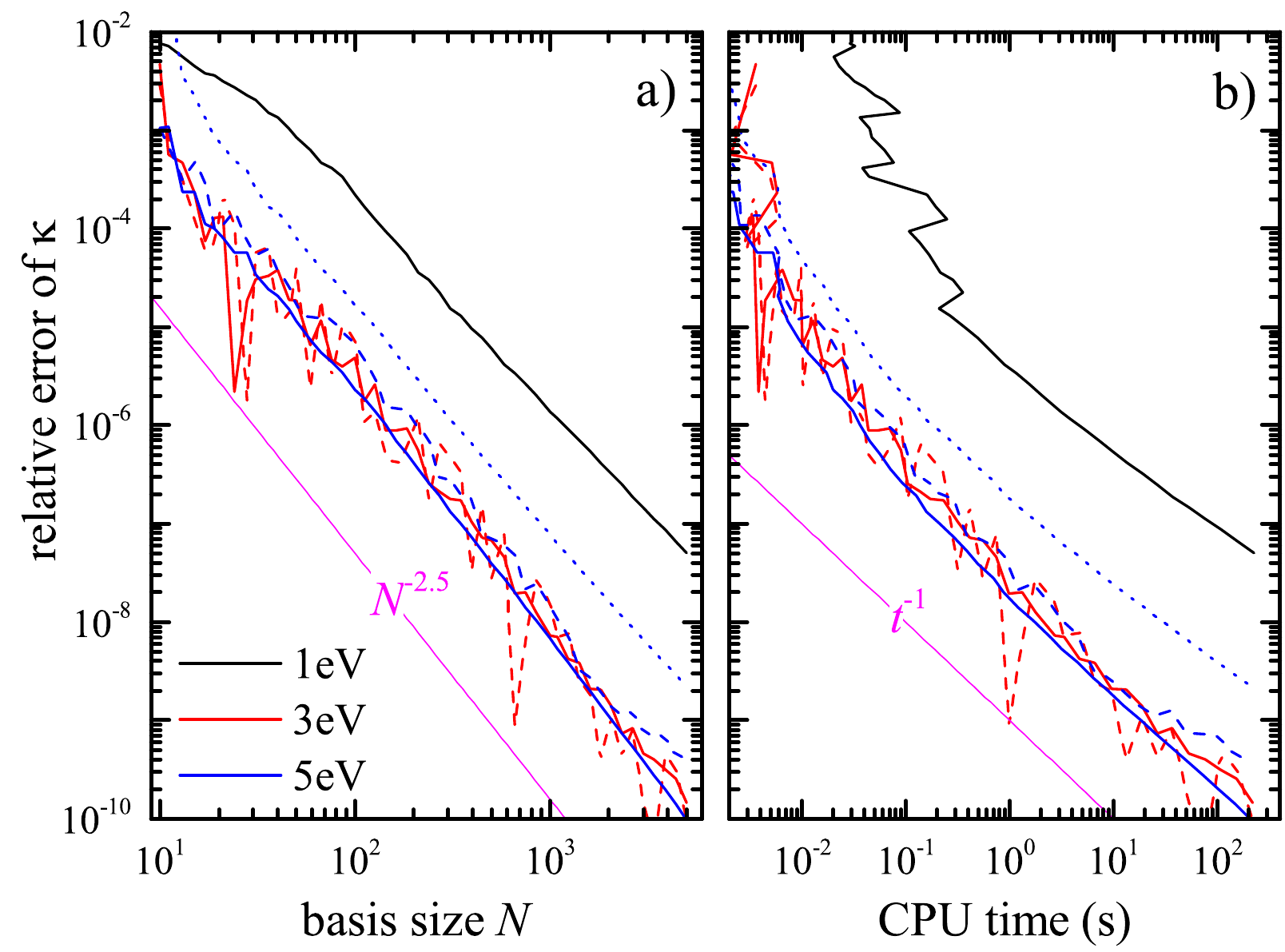}
	\caption{
		Relative error of the in-plane wavevector $\varkappa_j$ for the layer with the hole, for the WG modes ($j=1$:~solid, $j=2$:~dashed, $j=3$:~dotted) versus basis size $N$ (a) and CPU time (b), for different photon energies $\hbar\omega$ as indicated. The approximate convergence scalings $\propto N^{-2.5}$ and $\propto t^{-1}$ are indicated.}
	\label{fig:RSEConv}
\end{figure}

In this section, we present the convergence of the fixed-frequency RSE~\Eq{w-RSE2} for the slot region of the waveguide considered in the main text (see \Fig{fig:Spectra}(a)), versus basis size $N$. The relative error of the in-plane wavevector $\varkappa$ for the WG modes is shown in \Fig{fig:RSEConv} versus basis size $N$ and computational time on a CPU Intel Core i7-5830K. We find a convergence of the relative error scaling with $N^{-2.5}$, which is close to the $N^{-3}$ scaling of the RSE \cite{Muljarov10,Doost12}. The somewhat slower convergence can be related to the residual role on the continuum represented by the cut, which does not allow for a natural discretization.
In terms of computing time, the relative error scales approximately as $t^{-1}$ for large $N$, where it is dominated by the diagonalization of a non-sparse matrix with a computational complexity scaling as $N^3$.

\section{S-Matrix for a layered inhomogeneity of the waveguide}\label{sec:SMatrix}

Let us suppose that the non-uniform WG is represented by $L$ uniform regions, so that in each region $l$ defined as $z_{l-1}<z<z_l$, the permittivity $\varepsilon(x,z)=\varepsilon(x,z_l)$ is constant, where $l=1,2,\dots, L$, and $z_0=-\infty$ and $z_{L}=\infty$.
In this case, \Eq{d2An} has an exact analytic solution. To find it, we introduce, for each region $l$, a vector $\vec{A}_l(z)$   and a matrix $\hat{M}_l$ having elements
\bea
(\vec{A}_l)_n(z)&=&A_n(z)\,,
\\
(\hat{M}_l)_{nm} &=& p^2_n\delta_{nm} + \omega^2 V_{nm}(z_l)\,,
\label{MainEq}
\eea
respectively, given by the expansion coefficients $A_n(z)$ and $z$-independent matrix elements $V_{nm}(z_l)$. Then~\Eq{MainEq} becomes
\begin{equation}
	-\frac{d^2}{dz^2} \vec{A}_l(z) =\hat{M}_l \vec{A}_l(z).
\end{equation}
Its solution in region $l$ can be written as
\bea
	\!\!\!\! \vec{A}_l(z) &=& \hat{E}_l \left(
	e^{i\hat{K}_l (z-z_l)} \vec{b}^+_l +
	e^{-i\hat{K}_l (z-z_l)} \vec{b}^-_l \right),
\nonumber\\
 &=& \hat{E}_l \left(
	e^{i\hat{K}_l (z-z_{l-1})} \vec{d}^+_l +
	e^{-i\hat{K}_l (z-z_{l-1})} \vec{d}^-_l \right),
\label{b-equ}
\eea
where $\hat{K}_l$ and $\hat{E}_l$ are, respectively, a diagonal matrix of the eigenvalues and a matrix of the corresponding eigenvectors of
the eigenvalue problem~\Eq{w-RSE2} which can bewritten as
\begin{equation}
	\hat{M}_l \hat{E}_l = \hat{E}_l \hat{K}^2_l\,,
	\label{w-RSE}
\end{equation}
where the eigenvalues $\varkappa$ form the diagonal matrix $\hat{K}_l$ and the eigenvectors with components $c_m$ columns the matrix $\hat{E}_l$.

$\vec{b}^+_l$ and $\vec{b}^-_l$ (or $\vec{d}^+_l$ and $\vec{d}^-_l$) in \Eq{b-equ} are some constant vectors having the meaning of amplitudes of waves propagating, respectively, in the positive and negative direction of $z$. Maxwell's BCs provide relations between these amplitudes in neighboring layers:
\bea
\hat{E}_l \left( \vec{b}^+_l + \vec{b}^-_l \right) &=& \hat{E}_{l+1} \left( \vec{d}^+_{l+1} + \vec{d}^-_{l+1} \right), \\
\hat{E}_l \hat{K}_l \left( \vec{b}^+_l - \vec{b}^-_l \right) &=& \hat{E}_{l+1} \hat{K}_{l+1} \left( \vec{d}^+_{l+1} - \vec{d}^-_{l+1} \right)\,.
\eea
Using these relations and the S-matrix approach~\cite{KoPRB88}, we find the S-Matrix $\hat{S}$ of the whole system, which relates the incoming ($\vec{b}^+_{1}$ and $\vec{d}^-_{L}$) and outgoing ($\vec{b}^-_{1}$ and $\vec{d}^+_{L}$) amplitudes, in the left ($l=1$) and the right ($l=L$) layers just at their interfaces:
\begin{equation}
	\begin{pmatrix}
		\vec{b}^-_{1} \\
		\vec{d}^+_{L}
	\end{pmatrix} = \hat{S}
	\begin{pmatrix}
		\vec{b}^+_{1} \\
		\vec{d}^-_{L}
	\end{pmatrix}.\label{Eq:Smatrix_b}
\end{equation}
For the indexes $i$ and $j$ corresponding
to WG modes, one can also calculate~\cite{Tikhodeev02} the power S-Matrix $P_{ij}=|S_{ij}|^2$, which connects power fluxes in outgoing WG modes with incoming WG modes.

\begin{figure}[htb]
	\includegraphics[width=\linewidth]{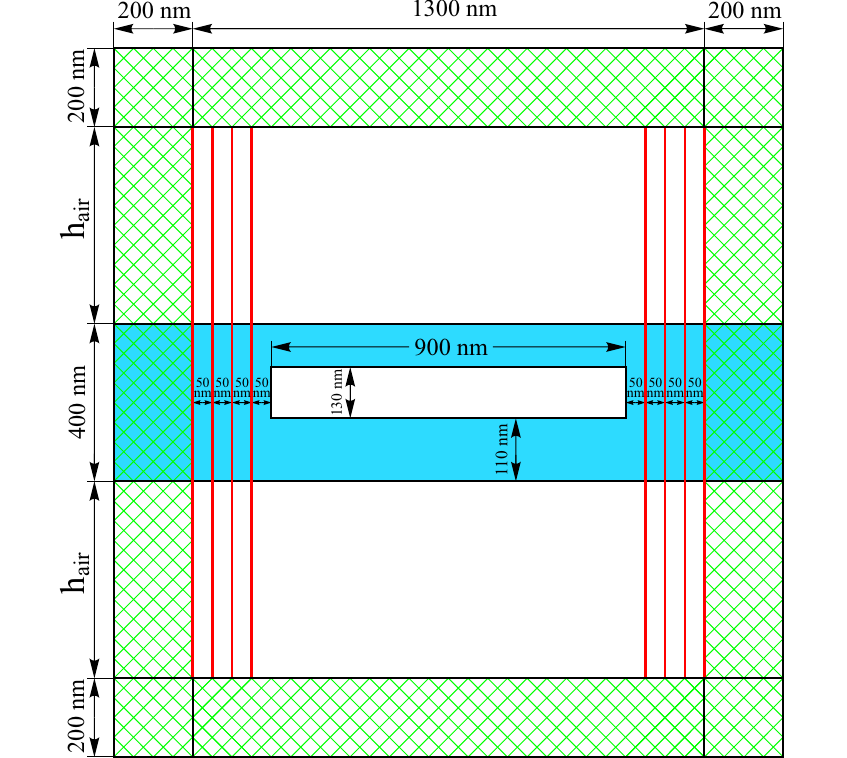}
	\caption{Schematic of the structure used in ComSol for \Fig{fig:Convergence} of the main text. Blue indicates the waveguide, green indicates PMLs, vertical red lines mark ports. We use one port per waveguide mode.
	}
	\label{fig:COMSOL}
\end{figure}

\section{a-FMM and ComSol models}
\label{sec:aFMMCOMSOL}
\subsection{a-FMM}
The details of the a-FMM method used for \Fig{fig:Convergence} of the main text is given in \Onlinecite{Pisarenco10}. We employed a quadratic PML with a damping strength $\sigma=2/|x-x_{t,b}|^p$ and $p=2$ (see Eq.\,(49) in Ref.~\Cite{Pisarenco10}). Here, $x_{t,b} = \pm (a + h_\mathrm{air})$ are the coordinates of the boundaries of the top and bottom PMLs, where $h_\mathrm{air}$ is the distance between the PML and the WG. To explore the convergence we increase the number of harmonics $N_g$ (see \Onlinecite{Pisarenco10}, where $N_g=2N+1$) and simultaneously increase the thickness $h_\mathrm{PML}/2$ of the PML and $h_\mathrm{air}$ according to
\begin{equation}
	h_\mathrm{PML}=h_\mathrm{air}=\frac{a}{2}\log N_g.
\end{equation}
This link between the parameters was found to be close to optimal for the convergence of the hole structure of \Sec{subsec:WGhole}, for the energies shown.

\subsection{ComSol}
The structure used in the ComSol calculations for \Fig{fig:Convergence} is shown in \Fig{fig:COMSOL}.
It has two variable parameters. The first one is the maximum element size of the mesh in air~$\Delta$.
The second variable parameter is the height of air slab~$h_\mathrm{air}$. We used
\be
\Delta = C_1 2^{-m} \quad \mbox{and} \quad h_\mathrm{air} = C_2 + C_3 m,
\ee
where $C_1=100$\,nm, $C_2=500$\,nm, $C_3=400$\,nm, and $m=0,\,0.1,\,0.2,\,0.3,\,\dots$ is parametrizing the convergence.
This link between the parameters was found to be close to optimal for the convergence of the hole structure of \Sec{subsec:WGhole}.

\end{document}